\documentclass[12pt]{article}
\usepackage{amssymb}
\usepackage[mathscr]{eucal}
\usepackage{url}
\usepackage{a4}
\usepackage{cite}
\usepackage{graphicx}
\usepackage{psfrag}
\usepackage{subfigure}
\usepackage{latexsym}
\usepackage{amsmath}

\newcounter{mnotecount}[section]

\begin{document}
\newcommand{\g}{$\bf \bar g g^{\alpha\beta}$}
\title{Celestial mechanics of elastic bodies II}
\author{Robert Beig\\Gravitational Physics\\Faculty of Physics, University
of Vienna\\ Boltzmanngasse 5, A-1090 Vienna, Austria\\[1cm]
Bernd G. Schmidt\\
Max-Planck-Institut f\"ur Gravitationsphysik\\ Albert-Einstein-Institut\\
Am M\"uhlenberg 1, D-14476 Golm, Germany}

\maketitle

\begin{abstract}
We construct time independent configurations describing a small elastic body moving in a circular orbit in the Schwarzschild spacetime. These configurations are relativistic versions of Newtonian solutions constructed by us previously. In the process we simplify and sharpen previous results of ours concerning elastic bodies in rigid rotation.

Keywords:  elastic bodies, rotation
\end{abstract}

\section{Introduction}
\label{introduction}
This work is part of a series of papers, where we construct time-independent solutions of the elastic equations in and without the presence of gravity, both in the Newtonian and relativistic context, see \cite{BS1,BS2,BS3,BS4}.
The present paper is basically a continuation of \cite{BS2}. Our main result concerns the rigid\footnote{We emphasize that 'rigid' here just refers to the nature of the motion. In other words, the body moves in the way the moon moves in the gravitational field of the earth - with the same side always pointing towards the earth. The body itself is by no means assumed to be rigid. Rather, for given material properties of the body, it is its deformation, from an initial stress-free state, under the combination of centrifugal and gravitational force acting on it, which is precisely our concern.}  motion of a small elastic body in a Schwarzschild spacetime. This is a relativistic analog of a result in \cite{BS2}, where we however considered the more unrealistic case of a - in principle arbitrary large - nonrelativistic elastic body which has certain discrete symmetries moving in a Newtonian $1/r$ - potential.

The configurations constructed in this series are always close to one where the elastic body is stressfree with no forces acting on it, i.e. where the body is at rest in a flat background spacetime. The technical tool is the implicit function theorem (IFT), and the main problem lies in the fact that the associated linear operator has nontrivial kernel and range. The method by which we deal with this problem differs from our previous works with the exception of \cite{BS4}. Namely we use the Liapunoff-Schmidt procedure (see e.g. \cite{GS}), where in the first step one finds solutions of a 'projected equation', in which the range of the linearized operator is forced to be onto, parameterized by a free element in the kernel. In the second, and crucial, step one views the condition that this family actually solves the full equation as an equation for this element in the kernel.
This finite dimensional condition, sometimes called the bifurcation equation, is also solved using the IFT. Here the associated linearized operator plays a central role in our work. This operator turns out to be given by a certain quadratic form $B$ on the Lie algebra of the Euclidean group.
The form $B$
depends only on the nonrelativistic limit of the forces acting on the body and the geometry of the body and can, for small bodies, be computed explicitly. When the body is small and sufficiently generic, i.e. its moments of inertia are pairwise different, this quadratic form has exactly the degeneracy coming from the symmetry of the problem. In the case of a body in circular motion in Schwarzschild, this symmetry is that given by the axial Killing vector $\frac{\partial}{\partial \phi}$. Thus the solution to our problem is only unique up to this isometry, and, in order to use the IFT, this has to be taken care of by suitably restricting the parameters on which the solution of the projected equation depends. \
As suggested by the above description our procedure, being based on the form $B$, is reasonably universal and can also be applied to other time-independent problems in nonrelativistic and relativistic elasticity. We point out some of these in the final section of this work, entitled 'Final Comments'.


\section{Elastodynamics and Elastostatics}
Let $(M,g_{\mu\nu})$ a relativistic spacetime, i.e. $M$ is a smooth 4-manifold with smooth Lorentz metric $g_{\mu\nu}$ of signature $(-+++)$, which is orientable and time-orientable. Elasticity theory deals with maps $f:M \rightarrow \mathcal{B}$, where $\mathcal{B}$, the 'material manifold' or body, will in our case be simply given as a domain $\subset (\mathbb{R}^3,\delta_{AB})$, flat Euclidean 3-space. The material manifold should heuristically be thought of as an abstract space of particles making up the elastic continuum. The maps $X^A = f^A(x^\mu)$ should be orientation-preserving and non-degenerate in the sense that $f^A{}_{,\mu}$ has everywhere 1-dimensional kernel. In addition this kernel should be timelike w.r. to $g_{\mu\nu}$. Thus, on the 'physical body' $f^{-1}(\mathcal{B})$, the vector field $u^\mu$, defined as the future-pointing element of this kernel with $u^\mu u^\nu g_{\mu\nu} = - 1$, is the 4-velocity of the particle world lines, which are in turn given as the inverse images under $f$ of points of $\mathcal{B}$.  The quantity $H^{AB} = f^A{}_{,\mu}
f^B{}_{,\nu} g^{\mu\nu}$, or rather $H^{AB} - \delta^{AB}$, should be thought of as 'strain'. Another basic object on $f^{-1}(\mathcal{B})$ is the number density $n$ of the material defined by the equation
\begin{equation} \label{x}
f^A{}_{,\mu}(x) f^B{}_{,\nu}(x) f^C{}_{,\sigma}(x) \,\epsilon_{ABC} = n(x) \,\epsilon_{\mu\nu\sigma\tau}(x) \,u^\tau(x)
\end{equation}
Because of the orientation-preserving property of $f$ here holds $n > 0$. The material properties are encoded in a function $\rho(f^A, H^{BC})$.
This function has two meanings. It plays the role of the Lagrangian density of elasticity. Secondly it is the energy density of the material in its own rest frame. To see this one computes the canonical energy momentum tensor $\mathcal{T}$ - which is minus the metric energy momentum tensor $T_\mu{}^\nu$ (see \cite{BS3}) - with the result that
\begin{equation}
\mathcal{T}_\mu{}^\nu = \frac{\partial \rho}{\partial f^A{}_{,\nu}} f^A{}_{,\mu}  - \rho \,\delta_\mu{}^\nu = - \rho\, u_\mu u^\nu + n \,\sigma_{AB} f^A{}_{,\mu} f^B{}_{,\rho}\,g^{\nu\rho}\,,
\end{equation}
where the 'second Piola stress' $\sigma_{AB}$ is given by
\begin{equation}
- \,\sigma_{AB} = 2 \frac{\partial \epsilon}{\partial H^{AB}}\hspace{0.4cm}\mathrm{with}\,\,\rho = n \,\epsilon\,.
\end{equation}
Note that quantities like $\rho$, $n$ or $u^\mu$ can be viewed as functions of $(f^A, H^{BC})$ or, like in (\ref{x}), as fields on spacetime, given a configuration $f^A(x)$. We follow the custom in Lagrangian field theory of expecting the reader to infer the correct interpretation from the context.
The field equations, given by $- \,\mathcal{E}_A = \partial_\nu (\frac{\partial \rho}{\partial f^A{}_{,\nu}}\,\sqrt{-g}) - \frac{\partial{\rho}}{\partial f^A} \sqrt{-g} = 0$, can be shown \cite{BS3} to be equivalent to $\nabla_\nu T_\mu{}^\nu = 0$.
They form a system of quasilinear equations for the unknown $f$. As for the nature of these equations, we shall in this work be concerned with configurations near one $\bar{f}$, which is both strainfree, i.e. $\bar{H}^{AB} = \delta^{AB}$ and stressfree, i.e. $\bar{\sigma}_{AB} = 0$. The leading-order derivative term of $\mathcal{E}_A$ has the form $M^{\mu\nu}_{AB} f^B{}_{,\mu\nu}$ where, for a stressfree state $\bar{f}$ \cite{BS3},
\begin{equation}
\bar{M}^{\mu\nu}_{AB} = \bar{\rho}\, \delta_{AB}\,\bar{u}^\mu \bar{u}^\nu -
2 \bar{n} \bar{L}_{ACBD} \bar{f}^C{}_{,\sigma} \bar{f}^D{}_{,\tau}g^{\mu \sigma} g^{\nu \tau}
\end{equation}
Here $\bar{L}_{ACBD} = \frac{\partial^2 \epsilon}{\partial H^{AC} \partial H^{BD}}|_{f = \bar{f}}$. When $\bar{\rho} > 0$ and under suitable positivity hypotheses on $\bar{L}_{ACBD}$ this implies hyperbolicity of the 2nd order operator $\mathcal{E}_A$ \cite{B}. The boundary conditions we will require in this work are given by
\begin{equation}
T_\mu{}^\nu n_\nu|_{f^{-1}(\partial \mathcal{B})} = 0 \Leftrightarrow \,\,u^\mu n_\mu|_{f^{-1}(\partial \mathcal{B})} = 0\,,\,\sigma_{AB}f^B{}_{,\mu} n^\mu |_{f^{-1}(\partial \mathcal{B})} = 0
\end{equation}
The first of these simply expresses the fact that there should be no matter outside $\overline{f^{-1}(\mathcal{B})}$ and the second is the condition of vanishing normal stress, which expresses the fact that we are dealing with a freely floating body that is to say in our case a body subject to no external forces other than gravity. If matter is coupled to the Einstein equations (which it is not in the present work), these conditions at the interface of matter and vacuum are mandatory anyway.\\
We will in this paper be concerned with solutions of the elastic equations where $u^\mu$ is proportional to a timelike Killing vector field $\zeta^\mu$ of the background metric $g$, thus turning the hyperbolic equations into elliptic ones.\\
The reduction turns the action $S[f;g] = \int_{f^{-1}(\mathcal{B})} \rho (f^A, H^{BC}) \sqrt{- \mathrm{det}g} \,d^4x$ into
\begin{equation}\label{redact}
S[\bar{f};V,h] = \int_{\bar{f}^{-1}(\mathcal{B})} \rho (\bar{f}^A, \bar{H}^{BC}) V \sqrt{\mathrm{det}h} \,d^3x\,,
\end{equation}
where $\bar{f} = N \rightarrow \mathcal{B}$ with $\bar{f} = f \circ \pi^{-1}$ with $\pi$ the projection of $M$ to the quotient space $N$ of $M$ under the action of $\zeta$, $\mathrm{det} g = - V^2 \,\mathrm{det} h$ with $- V^2 = g_{\mu\nu} \zeta^\mu \zeta^\nu$, $h_{ij}$ the natural (quotient) metric on $N$ and $\bar{H}^{AB} = \bar{f}^A{}_{,i} \bar{f}^B{}_{,j}\,h^{ij}$. By abuse of notation we omit the the bar in $\bar{f}$ in what follows. \\
The Euler-Lagrange equations for the action (\ref{redact}) plus boundary conditions are given by
\begin{equation} \label{ELs}
D_j (V\sigma_i{}^j) = n \epsilon D_iV\,\,\,\,\mathrm{in}\, f^{-1}(\mathcal{B}),\hspace{0.5cm}\sigma_{ij}n^j|_{f^{-1}(\partial \mathcal{B})} = 0
\end{equation}
where $\sigma_{ij} = - 2 \,n f^A{}_{,i}f^B{}_{,j} \,\frac{\partial \epsilon}{\partial H^{AB}}$ and
$D_i$ denotes covariant derivative w.r. to the metric $h_{ij}$.
This is a quasilinear system of 2nd-order PDE's (which, under appropriate conditions on $\rho$ - see Sect.3 - are elliptic)
with Neumann-type boundary conditions. The fact that these latter conditions are imposed on an a priori unknown boundary, makes it convenient - if not necessary - to go over to the material picture in which the unknown $f$ is replace by the unknown $\Phi = f^{-1}$. The new field equations can then either be inferred from (\ref{ELs}) using standard identities (see e.g. \cite{MH}). Alternatively one can substitute $f$ by $\Phi$ in the action (\ref{redact}), rendering the material action
\begin{equation}\label{matact}
S'[\Phi;V,h] = \int_\mathcal{B} \epsilon (X^A, H^{BC})\,V \circ \Phi\, d^3 X\,.
\end{equation}
Here $H^{AB}$ has to be understood as $H^{AB} = \Psi^A{}_i \Psi^B{}_j (h^{ij}\circ\Phi)$, where $\Psi^A{}_i \Phi^i{}_{,B} = \delta^A{}_B$. One can then vary (\ref{matact}) w.r. to $\Phi$.
Either way, the result can be written as
\begin{equation} \label{ELm}
\nabla_A [(V \circ \Phi)\,\sigma_i{}^A] = \epsilon (D_i V) \circ \Phi\,\,\,\,\mathrm{in}\,\,\mathcal{B},\,\,\,\,\,\sigma_i{}^A n_A|_{\partial \mathcal{B}} = 0
\end{equation}
where
\begin{equation}\label{sigma}
\sigma_i{}^A = \frac{\partial \epsilon}{\partial \Phi^i{}_{,A}} = - 2 \,(H^{AB} \Psi^C{}_{i} \,\frac{\partial \epsilon}{\partial H^{BC}}) \circ \Phi.
\end{equation}
Furthermore $\nabla_A \sigma_i{}^A = \partial_A \sigma_i{}^A -
(\Gamma_{\!ij}^k [h] \circ \Phi) \,\Phi^j{}_{,A} \,\sigma_k{}^A$. Note the identity
\begin{equation} \label{genid}
\int_\mathcal{B} (\xi^i \circ \Phi) \,\nabla_A [(V \circ \Phi)\,\sigma_i{}^A]\, dX - \int_{\partial \mathcal{B}}(\xi^i V \circ \Phi)\,\sigma_i{}^A n_A\, dS = 0
\end{equation}
for all Killing vectors $\xi^i$ of $(N,h_{ij})$. The identity (\ref{genid}) follows from the Gauss theorem together with
(\ref{sigma}) and the Killing equation $\mathcal{L}_\xi h = 0$. Relation (\ref{genid}) implies for all $\Phi$'s solving (\ref{ELm}) that
\begin{equation}\label{V}
\int_\mathcal{B} (\epsilon \,\mathcal{L}_\xi V) \circ \Phi\, dX = 0
\end{equation}
\section{The Schwarzschild case}
We now apply this scheme to the Schwarzschild spacetime $M$ of mass $m$, i.e.
\begin{equation}
g = -\left(1 - \frac{2m}{r}\right) dt^2 + \left(1 - \frac{2m}{r}\right)^{-1} dr^2 + r^2 (d\Theta^2 + \sin^2 \Theta d\phi^2),\,\,\,\,0<2m<r
\end{equation}
with Killing vector $\zeta$ given by $\xi = \partial_t + \Omega\,\partial_\phi$. Choosing as we do here a helical Killing vector for the reduction corresponds to considering elastic bodies in rigid circular motion. The norm of $\zeta$ is given by
\begin{equation}
V^2 = - \zeta^\mu \zeta_\mu = 1 - 2\left(\frac{m}{r} + \frac{\Omega^2 r^2 \sin ^2 \Theta}{2}\right)
\end{equation}
In order for $\zeta$ to be timelike, we have to restrict the range of $r$ to $\Omega^2 r^2 < 1 - \frac{2m}{r}$, which is empty for $27 \,\Omega^2 m^2 > 1$. Assuming that $27 \,\Omega^2 m^2$ is less than one\footnote{For an idea of orders of magnitude, this number is of order $10^{-31}$ for the earth-moon system.}, it follows that $N$ is given by the spherical shell $\{(2m < r_1 < r < r_2 < \infty,\Theta,\phi)\}$, where the explicit form of $r_1, r_2$ does not concern us\footnote{To see this, observe that the cubic $P(r) = \Omega^2 r^3 - r + 2m$ is positive at $r=2m$ and at infinity and has exactly one negative minimum at $r = (\sqrt{3} |\Omega|)^{-1}$ if $3 \sqrt{3} |\Omega| m < 1$} . Furthermore $V$ has critical points in $N$ given by $(r = L :\!= (\frac{m}{\Omega^2})^\frac{1}{3}, \Theta = \frac{\pi}{2})$. Now every point of $V$ in $N$, as shown by a short calculation, gives rise to an orbit of $\xi$ in $M$ which is tangent to a geodesic. These, in the present case, are of course the standard circular geodesics of Schwarzschild. Our procedure will be to put a small elastic body near a critical point of $V$.\\
We will in the following work for small $\Omega$ and small $m$. More precisely we will set $\Omega^2 = \lambda = \frac{m}{L^3}$, where $L$ is a fixed number in the interval $(r_1,r_2)$, and take $\lambda$ small. One checks by implicit differentiation of $P(r) = \lambda r^3 - r + 2 \lambda L^3$ w.r. to $\lambda$, that, as $\lambda$ tends to zero,
$r_1$ decreases monotonically towards zero and $r_2$ increases monotonically towards $\infty$. Let $x^i$ be the standard Euclidean coordinates associated with $(r,\Theta,\phi)$. The quotient metric $h_{ij}$, whatever its explicit form, has the property that
\begin{equation}\label{kij}
h_{ij} (x) = \delta_{ij} + \lambda \,k_{ij}(\lambda;x)
\end{equation}
with $k_{ij}$ a smooth function of its arguments. Also
\begin{equation}
V (x) = 1 + \lambda \,v(x) + \lambda^2 v'(\lambda;x)\,,
\end{equation}
where
\begin{equation}\label{potential}
- \,v =   \frac{(x^1)^2 + (x^2)^2}{2} + \frac{L^3}{r}
\end{equation}
with $r = \sqrt{(x^1)^2 + (x^2)^2 + (x^3)^2}$. Notice that $v$ is just the rescaled Newtonian (gravitational-plus-centrifugal) potential.\\ \\
For the constitutive conditions we will write $\epsilon = 1 + W$, where $1$ corresponds to the rest mass contribution to $\epsilon$ (we set $c=1$) and $W$ plays the role of what non-relativistically is called stored-energy function. The function $W (\Phi,\partial \Phi)$ will henceforth be assumed to just be a function of $H^{AB}$ ('homogenous material'). This is to merely simplify notation. Our results will still be valid when $W = W(X^A, H^{BC})$ or equivalently
$W = W(f^A, H^{BC})$ in spatial terms. Furthermore $W$ should satisfy:\\ \\
(i) \hspace{0.4cm}$W|_{H^{EF} = \,\delta^{EF}} \,= 0,\hspace{0.6cm}\frac{\partial W}{\partial H^{AB}}|_{H^{EF} =\,
\delta^{EF}} = 0$\\ \\
(ii)\hspace{0.4cm}$\bar{L}_{ABCD} = \frac{\partial^2 W}{\partial H^{AB} \partial H^{CD}}|_{H^{EF} = \,\delta^{EF}}$ is positive definite on symmetric 2-tensors on $\mathbb{R}^3$. This is an ellipticity condition.\\ \\
An important example, in fact the generic one in the case of isotropic materials, of a quadratic form $\bar{L}$ satisfying the 'pointwise stability condition' (ii) is
\begin{equation}
\bar{L}_{ABCD} = \mu_1 \,\delta_{AB}\delta_{CD} + 2 \mu_2\, \delta_{C(A}\delta_{B)D}\,,
\end{equation}
where $\mu_2 > 0$ and $3 \mu_1 + 2 \mu_2 > 0$.\\
We will take as reference solution a map $\Phi = \bar{\Phi}:\mathcal{B} \rightarrow (\mathbb{R}^3,\delta_{ij})$ which is an isometry, i.e. $\bar{H}^{AB} = \bar{\Psi}^A{}_i \bar{\Psi}^B{}_j \,\delta^{ij} = \delta^{AB}$. Thus, by (i) from above, it is also stressfree, i.e. $\bar{\sigma}_i{}^A$ is zero. Consequently $\bar{\Phi}$ is a solution of (\ref{ELm}) for $\lambda = 0$. We will seek a solution of (\ref{ELm}) for $\lambda > 0$ by perturbing around such an isometric solution. There will be no loss in assuming this isometry to be the identity map, i.e. $\bar{\Phi}^i (X) = \delta^i{}_A X^A$.\\
We will need expressions for our equations for small $\lambda$, but in an arbitrary state $\Phi$. For that define $\mathring{H}^{AB} = \Psi^A{}_i \Psi^B{}_j \,\delta^{ij}$ and write
\begin{equation}
H^{AB} = \mathring{H}^{AB} + \lambda \kappa^{AB}(\Phi,\partial \Phi) + \lambda^2 \kappa'^{AB}(\lambda;\Phi, \partial\Phi)
\end{equation}
and ($\mathring{\sigma}_i{}^A = \frac{\partial \mathring{W}}{\partial \Phi^i{}_{,A}}$, where $\mathring{W}(\partial \Phi) = W(\mathring{H}^{AB})$)
\begin{equation}
\sigma_i{}^A (\Phi,\partial \Phi) = \mathring{\sigma}_i{}^A (\partial \Phi) + \lambda\,\tau_i{}^A(\Phi,\partial\Phi) + \lambda^2 \,\tau'_i{}^A(\lambda;\Phi,\partial \Phi)
\end{equation}
We note for later use that
\begin{equation}
\mathring{\sigma}_i{}^A = - 2\, \frac{\partial \mathring{W}}{\partial \mathring{H}^{BC}} \mathring{H}^{CA} \Psi^B{}_i
\end{equation}
and
\begin{equation}\label{later}
\kappa^{AB} (\Phi, \partial \Phi) = \Psi^A{}_i \Psi^B{}_j \,l^{ij}(\Phi)\,,
\end{equation}
where $l^{ij}(\Phi) = - \delta^{il}\delta^{jm} k_{lm}(0;\Phi)$, where $k_{ij}$ is defined by (\ref{kij}).
One now finds that the field equation in (\ref{ELm}) can be written as ($\partial_i = \partial/\partial \Phi^i$)
\begin{multline}\label{order}
\partial_A \mathring{\sigma}_i{}^A - \lambda (1 + \mathring{W})\partial_i v \circ \Phi - \lambda\frac{\partial \mathring{W}}{\partial \mathring{H}^{AB}} \partial_i\kappa^{AB} = \\
- \lambda \,\partial_A \tau_i{}^A + \lambda^2 F_i(\lambda;\Phi,\partial \Phi)\hspace{0.4cm}\mathrm{in}\,\,\mathcal{B}
\end{multline}
and the boundary condition as
\begin{equation}\label{order1}
\mathring{\sigma}_i{}^A n_A |_{\partial\mathcal{B}} = - \lambda\,\tau_i{}^A n_A |_{\partial\mathcal{B}} + \lambda^2 f_i(\lambda;\Phi,\partial \Phi)|_{\partial \mathcal{B}}.
\end{equation}
Note that the last term on the left in (\ref{order}) comes from the Christoffel term in $\nabla_A \sigma_i{}^A$ in (Eq.\ref{ELm}). It is however more convenient to derive (\ref{order},\ref{order1}) by first expanding the Lagrangian density of the material action (\ref{matact})
\begin{equation}
L'(\Phi,\partial \Phi) = \epsilon\,\, (V\circ \Phi) = (1 + \lambda\, v)(1 + \mathring{W}) + \lambda \frac{\partial \mathring{W}}{\partial \mathring{H}^{AB}} \kappa^{AB} + O(\lambda^2)
\end{equation}
and inserting into
\begin{equation}\label{L'}
\partial_A \,\frac{\partial L'}{\partial \Phi^i{}_{,A}} - \frac{\partial L'}{\partial \Phi^i} = 0
\end{equation}
We now use identity (\ref{genid}) with $h_{ij}$ the flat metric and $V=1$. Then, using the boundary condition (\ref{order1}), terms of order $\lambda^0$ drop out. Thus a necessary condition on $\Phi$ in order to solve (\ref{order}) for $\lambda > 0$ is
\begin{multline}\label{necessary}
A_\xi[\Phi] = \int_\mathcal{B}
[-(1 + \mathring{W}) (\xi^i \partial_i v)\circ\Phi - \frac{\partial \mathring{W}}{\partial \mathring{H}^{AB}}\, (\xi^i \circ \Phi)\partial_i\kappa^{AB} -
\xi^i{}_{,j}\Phi^j{}_{,A}\,\tau_i{}^A -\\
- \lambda (\xi^i \circ \Phi)] F_i\,dX +
\lambda \int_{\partial \mathcal{B}} (\xi^i \circ \Phi) f_i \,dS = 0
\end{multline}
for any of the six independent Killing vectors $\xi$ of flat space.
The volume term in (\ref{necessary}) corresponds to total force and total torque acting on the integrand. For the third term in (\ref{necessary}) the quantity $\tau_i{}^A$ can be written as
\begin{equation}\label{above}
\tau_i{}^A = -2\left(\frac{\partial^2 \mathring{W}}{\partial \mathring{H}^{BC}\mathring{H}^{DE}}\mathring{H}^{EA} \Psi^D{}_i \kappa^{BC} - \frac{\partial \mathring{W}}{\partial \mathring{H}^{BE}} v \mathring{H}^{EA}+  \frac{\partial \mathring{W}}{\partial \mathring{H}^{BE}} \kappa^{CA} \Psi^B{}_i \right)
\end{equation}
The terms in (\ref{above}) involving $\mathring{H}^{EA}$ do not contribute to (\ref{necessary}) due to $\delta^{k(j}\xi^{i)}{}_{,k} = 0$.\\ \\
{\bf{Remark}}: When $(N,h_{ij},V)$  has a continuous symmetry, in the present case $\partial_\phi$, the relation (\ref{necessary}) holds independently of (\ref{order}), provided only that $\Phi$ satisfies the boundary condition
(\ref{order1}). To see this, first observe that the symmetry of the Lagrangian $L'$ under spatial axial rotations together with the fact that our boundary condition is 'natural' in that $\int_\mathcal{B} \frac{\partial L'}{\partial \Phi^i{}_{,A}}\, n_A dS = 0$ implies that the contraction of (\ref{L'}) whence that of (\ref{order}) with $\partial_\phi$ vanishes identically. Integrating over $\mathcal{B}$ and again using the boundary condition then implies that (\ref{necessary}) with $\xi = \partial_\phi$ holds independently of (\ref{order}).\\ \\
Setting $\lambda = 0$ and $\Phi = \bar{\Phi}$ in (\ref{necessary}), there follows
\begin{equation}\label{scaled}
\int_\mathcal{B} (\mathcal{L}_\xi v)\circ \bar{\Phi}\,dX = 0
\end{equation}
for all translational and rotational Killing vectors, which in turn means that the total force and total torque
of the potential $v$ acting on the physical body $\bar{\Phi}$ be zero. Note this is the same condition as the one arising in the nonrelativistic case. Moreover, as will be important later, the derivative of $A_\xi$ w.r. to $\Phi$ at $\Phi = \bar{\Phi}$ and $\lambda = 0$ in directions corresponding to Euclidean motions is given by the derivative of $\int_\mathcal{B} (\xi^i \partial_i v)\circ \Phi\,dX$ in those directions, namely (see also Appendix)
\begin{equation}\label{remarks}
B_v(\xi, \eta) = \int_\mathcal{B} (\mathcal{L}_\eta  \mathcal{L}_\xi v)\circ \bar{\Phi}\,dX
\end{equation}
The reason that the terms involving $W_0$ or $\frac{\partial \mathring{W}}{\partial \mathring{H}^{AB}} $ in (\ref{necessary}) at $\lambda = 0$ do not contribute is firstly they are clearly zero at $\Phi = \bar{\Phi}$. Secondly the one involving $\mathring{W}$ also has vanishing derivative at $\Phi = \bar{\Phi}$, and the ones involving
$\frac{\partial \mathring{W}}{\partial \mathring{H}^{AB}}$ have vanishing derivative at $\Phi = \bar{\Phi}$ in Killing directions $\xi$ since $\delta_\xi \mathring{H}^{AB} = 0$.
\section{Small generic bodies}
Our first task will be to choose a sufficiently generic, sufficiently small domain $\mathcal{B}$ near a critical point of $v \circ \bar{\Phi}$ such that (\ref{scaled}) holds. Such points are, to leading order in $\lambda$, critical points of $V$ and hence correspond to rigid circular motion in the Schwarzschild field. This will crucially depend on the degeneracy properties of the quadratic form $B_v(\xi,\eta)$ on the Lie algebra of the Euclidean group. The conditions we will pick will turn out to be sufficient also for the existence of solutions to the full elastic equations - both in the nonrelativistic and the relativistic case. \\
Following the procedure described in the Appendix, we first replace $v$ by a truncated $v$ called $v_\mathrm{tr}$, its Taylor expansion up to 2nd order around a critical point $x_0$
\begin{equation} \label{fij}
v_\mathrm{tr} = v(x_0) + \frac{1}{2}f_{ij} (x-x_0)^i(x-x_0)^j\hspace{0.8cm}f_{ij} = v_{,ij}(x_0)
\end{equation}
Recall that critical points $x_0$ have $(x_0{}^1)^2 + (x_0{}^2)^2 = L^2$ and $x_0{}^3 = 0$. For concreteness we will take $x_0 = (L,0,0) = \bar{\Phi}(X_0) = X_0$. We will also assume that $\mathcal{B}$, thus $\bar{\Phi}(\mathcal{B})$ lie outside the respective origins. This will be necessary in order for the slightly deformed body
 to lie inside $N$, which in the interior extends only up to some small radius $r_1$ depending on $\lambda$, as explained earlier. Define
\begin{equation} \label{regular}
\Theta^{ij} = \int_\mathcal{B}(\bar{\Phi}(X) - x_0)^i(\bar{\Phi}(X) - x_0)^j dX.
\end{equation}
This, modulo a trace-term, is the inertia tensor of
$\mathcal{B}$ (with density $\equiv 1$), centered at $x=x_0$ and w.r. to the map $\bar{\Phi}$. As described in the Appendix, we can arrange for the total force to be zero by performing a translation on $\mathcal{B}$ so that
the center-of-mass $D^i = |\mathcal{B}|^{-1}\int_\mathcal{B} \bar{\Phi}^i(X) dX$ is equal to $x_0^i$. Also, by a suitable rotation of $\mathcal{B}$, we can arrange for $f^i{}_j$ and $\Theta^i{}_j$ to commute which in turn sets the total torque $= 0$. Now we intend to show that, when $v_{\mathrm{tr}}$ is replaced with the full potential $v$, by yet another translation and rotation we can achieve the same goal for a suitably scaled version $\mathcal{B}_\rho = \{X \in \mathbb{R}^3|\frac{1}{\rho} (X - X_0) \in \mathcal{B} - X_0\}$ of $\mathcal{B} = \mathcal{B}_1$.
First one explicitly computes $B_v(\xi,\eta)$ for $v = v_\mathrm{tr}$. As basis for the Euclidean Killing vectors we choose translations $c^i \partial_i$ and rotations around $x_0$, i.e. $\epsilon^i{}_{jk} p^k (x - x_0)^j \partial_i$. It turns out that ($|\mathcal{B}|$ is the Euclidean volume of $\mathcal{B}$)
\begin{equation}
B_v(\xi,\xi') = |\mathcal{B}|f_{ij} c^i c'^j + \tau_{ij}p^i p'^j\,,
\end{equation}
where $\tau$, in a basis in which both $f$ and $\Theta$ are diagonal, is also diagonal and
\begin{equation}\label{basis}
\tau_{11} = (f_{33} - f_{22})(\Theta_{22}- \Theta_{33}),\tau_{22} = (f_{11} - f_{33})(\Theta_{33}- \Theta_{11}),\tau_{33} = (f_{22} - f_{11})(\Theta_{11}- \Theta_{22})
\end{equation}
Then we try satisfy the equilibrium conditions (\ref{scaled}) for the full potential $v = v_\mathrm{tr} + O((x-x_0)^2)$ for $\mathcal{B}_\rho$ by a small Euclidean motion. We thus seek to solve
\begin{equation}\label{epsilon}
M_{(c,p)}(\rho; \Phi) = \int_{\mathcal{B_\rho}}(\mathcal{L}_{\xi_{(c,p)}} v) \circ \Phi\,\,dX = 0\,
\end{equation}
for small $\rho > 0$ where, as in the Appendix, $\Phi$ is the map $\bar{\Phi}$ composed with an element of the Euclidean group on $(\mathbb{R}^3, \delta_{ij})$ close to the identity. Here $\xi_{(c,p)}$ is given by
\begin{equation}\label{basis1}
\xi^i_{(c,p)}(x) = \epsilon^i{}_{jk}p^k (x - x_0)^j + c^i\,,\hspace{0.8cm}c^i = \mathrm{const},\,\, p^i = \mathrm{const}
\end{equation}
The quantity $M$ in (\ref{epsilon}) should be viewed as a $\rho$-dependent function on $\mathbb{E}(3)$ taking values in $\mathbb{E}(3)^\star$. The map $\Phi$ can be written as $\Phi = \mathrm{exp} \,\xi_{(C,P)} \circ \bar{\Phi} = \Phi_{(C,P)}$. In order to be able to apply the IFT, we modify $M_{(c,p)}(\rho;\Phi_{(C,P)})$ as follows
\begin{equation}
\overline{M}_{(c,p)}(\rho;\Phi_{(C,P)}):= \rho^{-4}M_{(c,0)}(\rho;\Phi_{(\rho C,P)}) +  \rho^{-5}M_{(0,p)}(\rho;
\Phi_{(\rho C,P)})\,
\end{equation}
Noting that $\mathrm{exp}\,\xi_{(\rho c,p)}\rho \,x = \rho \,\mathrm{exp}\,\xi_{(c,p)}x$ and recalling that $\bar{\Phi}$ is the identity map, in particular linear, it follows that
\begin{equation}
\overline{M}_{(c,p)}(\rho;\Phi_{(C,P)}) = \int_\mathcal{B}(\mathcal{L}_{\xi_{(c,p)}} v_{\mathrm{tr}}) \circ \, \mathrm{exp}\,\xi_{(C,P)}\bar{\Phi}\,\,dX + O(\rho)
\end{equation}
Thus
\begin{equation}
\overline{M}_{(c,p)}(0;\bar{\Phi}) = 0
\end{equation}
and the linearization-at-the-identity $\delta_{(c',p')} = \partial_C|_{(C=0,P=0)}c' + \partial_P|_{(C=0,P=0)}p'$ of $\overline{M}$ at $\rho = 0$ is given by
\begin{equation}
\delta_{(c',p')} \overline{M}_{(c,p)}|_{\rho = 0} = B_{v_{\mathrm{tr}}}(\xi_{(c,p)}, \xi_{(c',p')})
\end{equation}
Now, for the potential $v$ given by (\ref{potential}), $x_0 = (L,0,0)$ and $f_{ij}$ from (\ref{fij}) is given by
\begin{equation}
f_{ij} = \mathrm{diag} (-3,0,1)\,,
\end{equation}
and thus $v_{\mathrm{tr}}$ is up to an irrelevant additive constant given by
\begin{equation}
- \,v_{\mathrm{tr}} = \frac{1}{2}\,[3 (x^1 - L)^2 - (x^3)^2]
\end{equation}
Thus, when the $\Theta$'s are pairwise unequal, the form $B_v$ has rank $5$, namely its null space is spanned by $\xi = \partial_2 = \xi_{((0,1,0),(0,0,0))}$, which is the remnant of the invariance of the full potential under
$\xi = \partial_\phi$, i.e. rotations around the $x^3$ - axis. Note that
\begin{equation}
\partial_\phi = \xi_{((0,L,0),(0,0,-1))}
\end{equation}
Pick a linear 5-dimensional subspace $\mathbb{E}'$ of $\mathbb{E}(3)$ transversal to $\partial_2$. For concreteness let us take for $\mathbb{E}'$ the set $\{\xi_{(c^1,0,c^3),(p^1,p^2,p^3)}\}$. Next try solving $\overline{M}_{(c,p)}(\rho;\Phi_{(C,P)}) = 0$, with both $c^2 = 0$ and $C^2 = 0$, for small positive $\rho$. Since the restriction of the quadratic form $B_v$ to $\mathbb{E}'$ is non-degenerate, there exists, by the IFT, a unique solution $(\bar{C},\bar{P})$ in $\mathbb{E}'$ near $(0,0)$, whence a unique solution $(C_0,P_0) = (\rho \bar{C},\bar{P})$ of
$M_{(c,p)}(\rho;\Phi_{(C,P)}) = 0$. But $M_{((0,L,0),(0,0,-1))}(\epsilon;\Phi_{(C_0,P_0)}) = 0$, Thus, by transversality of $\mathbb{E}'$ to $\xi_{((0,1,0),(0,0,0))}$, $M_{(c,p)}(\rho;(C_0,P_0)) = 0$ for all $(c,p) \in \mathbb{E}(3)$, and we are done.\\
Let $\bar{\Phi}_\rho = \mathrm{exp}\,\xi_{(C_0,P_0)} \bar{\Phi}$. Then
\begin{equation}
B^\rho_v((c,p),(c',p')) = \int_{\mathcal{B_\rho}}(\mathcal{L}_{\xi_{(c,p)}} \mathcal{L}_{\xi_{(c',p')}}v) \circ \bar{\Phi}_\rho\,\,dX
\end{equation}
is again symmetric in $((c,p),(c',p'))$ for small positive $\rho$. Clearly this quadratic form has rank at most $5$. In fact, making $\rho$ smaller if necessary, $\partial_\phi$ stays the only element in the null space. To see this, note that
\begin{equation}\label{l4}
B^\rho_v((c,0),(c',0)) = \rho^3 B_v((c,0),(c',0)) + O(\rho^4)
\end{equation}
\begin{equation}\label{l5}
B^\rho_v((c,0),(0,p')) = \rho^4 B_v((c,0),(0,p')) + O(\rho^5)
\end{equation}
\begin{equation}\label{l6}
B^\rho_v((0,p),(0,p')) = \rho^5 B_v((0,p),(0,p')) + O(\rho^6)\,,
\end{equation}
where we recall that $B$ on the r.h. side refers to the truncated $v_\mathrm{tr}$, which we know explicitly.
Let us denote by $(\bar{c},p)$ the set $(c^1,0,c^3,p^1,p^2,p^3)$. Now recall that the determinant of the $5 \times 5$ matrix given by
\begin{displaymath}
\left(\begin{array}{c|c}
B_v((\bar{c},0),(\bar{c}',0)) & B_v((\bar{c},0),(0,p')) \\
\hline
B_v((\bar{c},0),(0,p')) & B_v((0,p),(0,p'))
\end{array}\right)
\end{displaymath}
is nonzero. But then the determinant of the analogous matrix whose submatrices are given by the respective l.h. sides of
(\ref{l4},\ref{l5},\ref{l6}) is $\rho^{3+3+5+5+5}=\rho^{21}$ times the above nonzero determinant plus terms of $O(\rho^{22})$. This proves our assertion.\\
We henceforth assume that $\mathcal{B}_\rho$ is of a form just constructed and that its principal moments of inertia are all different. By slight abuse we call such $\mathcal{B}_\rho$ again $\mathcal{B}$.
We now turn to solving (\ref{order},\ref{order1}) in the present scenario, which we for the moment write as
\begin{equation}\label{f1}
\partial_A \mathring{\sigma}_i{}^A = b_i (\lambda;\Phi,\partial\Phi) \,\,\,\,\,\mathrm{in}\,\,\mathcal{B}
\end{equation}
\begin{equation}\label{f2}
\mathring{\sigma}_i{}^A n_A|_{\partial \mathcal{B}} = \tau_i(\lambda;\Phi,\partial\Phi)|_{\partial \mathcal{B}}
\end{equation}
Let now $\mathscr{C}$ be a neighbourhood of $\Phi = \bar{\Phi}$ in $W^{2,p}(\mathcal{B},\mathbb{R}^3),\;p > 3$,
small enough so that each $\Phi \in \mathscr{C}$
has a $C^1$ - inverse (see p.224 of \cite{CI}).
Next let $\mathscr{E}$ be the quasilinear second-order operator sending each map $\Phi \in \mathscr{C}$ to
\begin{equation}\label{defE}
\mathscr{E}_i [\Phi](X) = (\partial_A \mathring{\sigma}_i{}^A) (X)
\end{equation}
Denote $\sigma_i$ by $(\mathring{\sigma}_i{}^A n_A)|_{\partial
\mathcal{B}}$, where $n_A$ is the outward normal to $\partial \mathcal{B}$.
Let the ``load space'' $\mathscr{L}$ be defined by $\mathscr{L} =
W^{0,p}(\mathcal{B},\mathbb{R}^3) \times W^{1 -
1/p,p}(\partial\mathcal{B},\mathbb{R}^3)$. It is then well known (see
\cite{VA}), that the operator $E$ sending $\Phi \in \mathscr{C}$ to
the pair $(b_i = \mathscr{E}_i, \tau_i = \sigma_i) \in \mathscr{L}$ is
well defined and $C^1$. The operator $E$ can not be onto because of (\ref{genid}) or (\ref{necessary}), namely there has to hold
\begin{equation}\label{equil}
\int_{\mathcal{B}} (\xi^i \circ \Phi) \,b_i\, dX - \int_{\partial \mathcal{B}} (\xi^i \circ \Phi)\,\tau_i\, dS = 0
\end{equation}
It is also a known fact (see \cite{VA} or \cite{MH}) that the linearized operator $\delta E$ at $\Phi = \bar{\Phi}$ given by
$(\partial_A \,\delta \mathring{\sigma}_i{}^A, \delta \mathring{\sigma}_i{}^A n_A|_{\partial \mathcal{B}})$, where
\begin{equation}\label{deltasigma}
\delta \mathring{\sigma}_i{}^A [\delta \Phi] = 4 \,\delta^{AB}\delta^{DF}\bar{\Psi}^C{}_i \bar{\Psi}^E{}_j \,\bar{L}_{BCDE}\,\delta\Phi^j{}_{,F}\,,
\end{equation}
viewed as an operator form $W^{2,p}(\mathcal{B},\mathbb{R}^3)$ into $\mathcal{L}$, has a 6-dimensional kernel given by $\delta \Phi^i = \xi^i \circ \bar{\Phi}$, where $\xi^i$ is a Euclidean Killing vector and range given by elements of $\mathcal{L}$ satisfying (\ref{equil}) with $\Phi = \bar{\Phi}$. Note for the kernel property the pointwise stability condition (ii) is vital. \\
We next turn to solving (\ref{f1},\ref{f2}) using a variant of the method of Liapunoff-Schmidt (see e.g. \cite{GS}). For this purpose we first define a linear projection operator $\mathbb{P}:(b,\tau) \in \mathcal{L} \mapsto (b'(b,\tau),\tau) \in \mathcal{L}$ as follows. Consider the 6-dimensional subspace of $\mathcal{L}$ given by $\xi_i = \delta_{ij} \,\xi^j$ with $\xi^i$ Killing vector fields on $\mathbb{R}^3$ in the first slot and $0$ in the second, i.e. $(\xi_i \circ \bar{\Phi},0)$. Since the bilinear form on Killing vectors given by the $L^2$ - inner product on $\mathcal{B}$ is non-degenerate\footnote{Namely, this quadratic form, say
$Q(\xi,\xi') = \int_{\mathcal{B}} (\xi^i \circ \bar{\Phi})(\xi'^j \circ \bar{\Phi})\delta_{ij}\, dX$, is in the basis (\ref{basis1}) given by $Q(\xi,\xi') = |\mathcal{B}|\, \delta_{ij}\, c^i c'^j + (\mathrm{tr} \Theta \,\delta_{ij} - \Theta_{ij})\, p^i p'^j$. Thus, since $\Theta_{ij}$ is positive definite, $Q$ is non-degenerate.}, this subspace is transversal to the subspace $\mathcal{L}' \subset \mathcal{L}$ given by elements of $\mathcal{L}$ satisfying (\ref{equil})\footnote{These could be called 'loads equilibrated at the identity'} with $\Phi = \bar{\Phi}$. Now $\mathbb{P}$ is defined as the unique operator on $\mathcal{L}$
projecting onto that subspace along Killing vectors as above.\\
Next let $\mathcal{C}'$ be the Banach subspace of $\mathcal{C}$ given $\Phi^2(x_0) = (x_0)^2 = 0$. The idea now is to solve a projected version of (\ref{f1},\ref{f2}), but, as it were, with an arbitrary element in the kernel of $\delta E'$, where the arbitrariness refers just to Killing directions transversal to $\partial_\Phi$. For this purpose consider the map $F: \mathbb{R}_{\geq 0} \oplus \mathbb{R}^5 \oplus \mathcal{C}' \rightarrow \mathcal{L}' \oplus \mathbb{R}^5$ given by
\begin{equation}
F:(\lambda,c^1,c^3,p^1,p^2,p^3;\Phi) \in \mathbb{R}_{\geq 0} \oplus \mathbb{R}^5 \oplus \mathcal{C}' \mapsto (b',\tau',d^{1},d^{3},q^{1},q^{2},q^{3}) \in \mathcal{L}' \oplus \mathbb{R}^5
\end{equation}
where
\begin{equation}
b'_i = \mathbb{P}(\mathcal{E}_i [\Phi] - b_i [\lambda;\Phi]),\,\tau'_i = \sigma_i [\Phi]- \tau_i [\lambda;\Phi]
\end{equation}
and
\begin{align} \label{and}
d^{1} = c^1 - \Phi^1(X_0) + L,\,\,\,d^{3} = c^3 - \Phi^3 (X_0),\;\;\;\;\;\;\;\;\;\;\;\;\;\;\;\;\nonumber\\
q^{1} = p^1 - \Phi^{[2}{}_{,3]}(X_0),\,\,\,q^{2} = p^2 -  \Phi^{[3}{}_{,1]}(X_0),\,\,\,q^{3} = p^3 -  \Phi^{[1}{}_{,2]}(X_0)
\end{align}
The numbers $(\lambda,c^1,c^3,p^1,p^2,p^3)$ should be viewed as parameters. It is clear from the above that $\Phi = \bar{\Phi} = \mathrm{id}$ solves the equation $F = 0$ for zero value of the parameters. \\
{\bf{Theorem 1}}: For small values of the 5+1= 6 parameters the equation $F = 0$ has a unique solution $\Phi$ near $\bar{\Phi}$.\\
{\bf{Proof}}: The operator $F$ is $C^1$ in all its variables. For the dependence on $\Phi$ this follows from arguments (see \cite{VA} and the Appendix in \cite{BSR}), based on the quasi-linear character of the operator $E$. The $C^1$ - dependence on the full set of variables is then straightforward to check. We now look at the linearized operator at $\bar{\Phi}$. Any $\delta \Phi$ in the tangent space at $\bar{\Phi}$ in $\mathcal{C}'$ satisfies the constraint
$\delta \Phi^2 (X_0) = 0$. The linearized operator is given by the 7-tuple:
\begin{equation}
\delta\Phi \mapsto (\partial_A \,\delta \mathring{\sigma}_i{}^A, \delta \mathring{\sigma}_i{}^A n_A|_{\partial \mathcal{B}},\delta\Phi^1(X_0),\delta\Phi^3 (X_0),\delta\Phi^{[2}{}_{,3]}(X_0),\delta\Phi^{[3}{}_{,1]}(X_0),
\delta\Phi^{[1}{}_{,2]}(X_0))
\end{equation}
where $\delta \mathring{\sigma}_i{}^A$ is given by (\ref{deltasigma}). Now the nullspace for the operator given by the first 2 slots are Killing vectors with $\delta \Phi^2(X_0) = 0$, which however have to be zero in order for being in the null space of the remaining slots. Thus our linearized operator has trivial kernel. Furthermore, for the first 2 slots being any equilibrated load, there always is a $\delta \Phi$ mapped into it. But we are free to add any Killing vector to $\delta \Phi$ with $\delta \Phi^2(X_0) = 0$ - and those exhaust the remaining slots. Thus the linearized operator has full range, and the theorem is proven using the IFT.\\
Now let $\Phi_{(\lambda;\bar{c},p)}$ with $(\bar{c},p) = (\bar{c}^1,\bar{c}^3,p^1,p^2,p^3)$ be the solution, near $0$ in $\mathbb{R}_{\geq 0} \otimes \mathbb{R}^5$, afforded by the above theorem. When $(\lambda;\bar{c},p)=0$, this solution clearly satisfies
the 6 conditions given by (\ref{necessary}).
We now show\\
{\bf{Theorem 2}}: We can, for sufficiently small $\lambda > 0$, find unique $(\bar{c},p)(\lambda)$, such that the 6 conditions (\ref{necessary}) on $\Phi_{(\lambda;\bar{c},p)}$ are valid.\\
{\bf{Proof}}: We first compute the derivative of $\Phi_{(\lambda;\bar{c},p)}$ w.r. to $(\bar{c},p)$ at $(\bar{c},p)=0$. Namely the quantity $\delta_{(\bar{c}',p')} \Phi = (\bar{c}'^\alpha \frac{d}{d \bar{c}^\alpha}|_{(0;0,0)}
+ p'^i \frac{d}{d p^i}|_{(0;0,0)}) \Phi_{(\lambda;\bar{d},p)}$, by implicit differentiation of (\ref{and}), satisfies the equations
\begin{align}
\delta_{(\bar{c}',p')} \Phi^1(X_0) = \bar{c}'^1,\,\,\,\,\delta_{(\bar{c}',p')} \Phi^3 (X_0) = \bar{c}'^3,\;\;\;\;\;\;\;\;\;\;\;\;\;\;\;\;\nonumber\\
\delta_{(\bar{c}',p')}\Phi^{[2}{}_{,3]}(X_0)=p'^1,\,\,\, \delta_{(\bar{c}',p')}\Phi^{[3}{}_{,1]}(X_0)=p'^2,\,\,\,
\delta_{(\bar{c}',p')} \Phi^{[1}{}_{,2]}(X_0)=p'^3
\end{align}
Furthermore $\delta_{(\bar{c}',p')} \Phi$ has to be an element of the null space of the operator $E$. Thus
\begin{equation}
\delta_{(\bar{c}',p')} \Phi = \xi_{(\bar{c}'^1,0,\bar{c}'^3,p'^1,p'^2,p'^3)}
\end{equation}
Next note that, by our choice of projection $\mathbb{P}$, the fields $\Phi_{(\lambda;\bar{c},p)}$ already satisfy the correct boundary condition (\ref{order1}). Thus the equilibration condition to be satisfied is (\ref{necessary}). So consider $A_\xi[\Phi_{(\lambda;\bar{c},p)}]$ in (\ref{necessary}) with $\xi \in \mathbb{E}'$, which vanishes
for $(\lambda;\bar{c},p) = (0;0,0)$. But, from the above computation combined with the remarks following (\ref{remarks}), it follows that, at $\lambda = 0$, the differential of $A_\xi$ w.r. to $(\bar{c},p)$ at $(\bar{c},p) = 0$ is nothing but the form $B_v$ restricted to $\mathbb{E}'$ - thus non-degenerate. It thus follows from the finite-dimensional IFT we can find unique $(\bar{c}(\lambda),p(\lambda))$ for small $\lambda > 0$, such that
$A_\xi[\Phi_{(\lambda;\bar{c}(\lambda),p(\lambda))}] = 0$ with $\xi \in \mathbb{E}'$. Finally note that $A_{\partial_\phi} [\Phi]$ vanishes for all values of $(\bar{c},p)$. This follows from the remark after (\ref{above}) together with the fact, again, that our choice of projection leaves the boundary condition untouched. So, again using the transversality of $\partial_\phi$ and
$\mathbb{E}'$ the result follows.\\
It is now easy to prove our main result, namely\\
{\bf{Theorem 3}}: The equations (\ref{ELm}) have a solution $\Phi_\lambda$ for small $\lambda > 0$. This is unique
up $\phi$-rotations of the form $(\Phi^1,\Phi^2,\Phi^3) \mapsto (\cos\!\phi\, \,\Phi^1 + \sin\! \phi \,\,\Phi^2,
-\sin\! \phi \,\,\Phi^1 + \cos\! \phi \,\,\Phi^2,\Phi^3)$.\\
{\bf{Proof}}: The combination of Thm.1 and Thm.2 provides us with a solution of
\begin{equation} \label{ELmproj}
\mathbb{P}\{\nabla_A [(V \circ \Phi_\lambda)\,\sigma_i{}^A] - \epsilon (D_i V) \circ \Phi_\lambda\} = 0\,\,\,\,\mathrm{in}\,\,\mathcal{B},\,\,\,\,\,\sigma_i{}^A n_A|_{\partial \mathcal{B}} = 0
\end{equation}
together with
\begin{equation}\label{contra}
\int_\mathcal{B} (\xi^i \circ \Phi_\lambda) \{\nabla_A [(V \circ \Phi_\lambda)\,\sigma_i{}^A] - \epsilon (D_i V) \circ \Phi_\lambda\} \, dX = 0\,\,\,\,\mathrm{for \,all}\, \xi\, \mathrm{in}\, \mathbb{E}(3)
\end{equation}
From the first of (\ref{ELmproj}) and the construction of $\mathbb{P}$ it follows that $\nabla_A [(V \circ \Phi_\lambda)\,\sigma_i{}^A] - \epsilon (D_i V) \circ \Phi_\lambda$ is a constant linear combination of Killing vectors $\xi_i \circ \bar{\Phi}$. Since $Q(\xi,\xi') = \int_{\mathcal{B}} (\xi^i \circ \bar{\Phi}) \,(\xi'^j \circ \bar{\Phi})\,\delta_{ij}\, dX$ is non-degenerate in the space of Killing vectors, so is
$\int_{\mathcal{B}} (\xi^i \circ \bar{\Phi}) \,(\xi'^j \circ \bar{\Phi})\,\delta_{ij}\, dX$, by continuity. That however contradicts (\ref{contra}) except when $\nabla_A [(V \circ \Phi_\lambda)\,\sigma_i{}^A] - \epsilon (D_i V) \circ \Phi_\lambda$ is zero. So we have in fact found a solution to the full, i.e. unprojected equations. This is unique in the class of $\Phi$'s obeying $\Phi^2(X_0) = 0$, which ends the proof of Thm.3.\\
\section{Final Comments}
In this paper we solved a class of time-independent problems in elasticity of the form given in (\ref{f1},\ref{f2}) near some stressfree state for $\lambda = 0$, using the IFT. The problem to be surmounted is that the linearized operator $E$ is not an isomorphism. Namely it has a 6-dimensional kernel which comes from the symmetry of the nonrelativistic elasticity operator (the l.h. side of (\ref{f2},\ref{f2})) under Euclidean motions. It also has a 6-dimensional co-kernel corresponding to the linearized version of the fact that force densities acting on an elastic body at rest have to satisfy the 6 conditions of being 'equilibrated' in the sense that total force and total torque acting on the body be zero. Our way of dealing with this problem is by, in the first step, solving a version of these equations which is projected on the range of this linear operator, with some free parameters corresponding to the nontrivial kernel of the linearized operator. In the next and crucial step one tries to fix these parameters in order so that the solution found in the previous step does indeed satisfy the 6 equilibration conditions. Actually the situation is a little more complicated than this: the r.h. sides of (\ref{f1},\ref{f2}) can, and in our case do, have a symmetry, namely spatial axial symmetry. This means on one hand there has to be a lack of uniqueness for the final solution. Dually, the equilibration condition corresponding to the axial Killing vector, i.e. the vanishing of total torque w.r. respect to the z-axis, is valid identically. So we need to fix the $\phi$ - rotations by hand and try solve the remaining 5 equilibration conditions for the remaining 5 parameters in the previous solution. This is possible, provided that the moments of inertia of the undeformed body are sufficiently generic, as shown by our analysis of the quadratic form $B_v$ on the Lie algebra of the Euclidean group which plays the role of the linearized operator of the equilibration conditions\footnote{We take this opportunity to point out that in our paper \cite{BS4} the computation of this linearized operator, i.e. (4.23) through (4.26) of that paper was wrong so that we missed imposing the condition that the inertia moments $\Theta_i$ satisfy $\Theta_1 \neq \Theta_3$ and $\Theta_2 \neq \Theta_3$.}. Moreover this quadratic form governs the solubility of the problem both in the relativistic and nonrelativistic case despite the fact that already the linearized solutions in both cases are different. \\
The method as outlined above can be applied to elastic problems other than the helical motion of small bodies in the Schwarzschild spacetime. One can for example redo the simpler problem in \cite{BSR} of a body in rigid rotation (the symmetry group in that case being $\mathbb{U}(1) \times \mathbb{R}$ i.e. cylindrical symmetry), using exactly the same method. This significantly simplifies the treatment of \cite{BSR} and, in addition, shows that a condition on the elastic constants imposed in the relativistic case is superfluous. One can also, as done for the nonrelativistic case of helical motion in Schwarzschild in \cite{BS2}, replace the smallness assumption on the body by imposing mirror symmetry with respect to the $X_2 = 0$ and $X_3 = 0$ - planes\footnote{The condition of a homogenous and isotropic elastic body imposed in the present paper for convenience then becomes essential}. This assumption eliminates 4 from the 5 parameters present in the equilibration condition has been used in the preprint \cite{Br} which also contains explicit calculations of linearized solutions. However the existence proof there is incomplete.\\ \\
{\bf{Acknowledgment}}: R.Beig was partially supported by Fonds zur F\"orderung der Wissenschaftlichen Forschung Proj.Nr. P20414-N16. He also acknowledges many discussions with S.Broda in connection with his preprint \cite{Br}.
\appendix
\section{The form $B$}
Let $\bar{\Phi}$ be an isometry $\bar{\Phi}:(\mathcal{B}, \delta_{AB}) \rightarrow (\mathbb{R}^3, \delta_{ij})$ with the property that
\begin{equation}\label{critical}
\int_\mathcal{B} (\mathcal{L}_\xi v)\circ \bar{\Phi}\,dX = 0
\end{equation}
for all Euclidean Killing vectors $\xi$. Next consider the functional
\begin{equation}\label{euact}
S[\Phi] = \int_\mathcal{B} v \circ \Phi \,dX\,,
\end{equation}
where $\Phi$ varies among all Euclidean motions of $(\mathbb{R}^3,\delta_{ij})$ composed with $\bar{\Phi}$. Clearly $S$ has a critical point at $\Phi = \bar{\Phi}$. Now the quadratic form $B$ on $\mathbb{E}(3)$ given in (\ref{remarks}), namely
\begin{equation}
B_v(\xi, \eta) = \int_\mathcal{B} (\mathcal{L}_\eta  \mathcal{L}_\xi v)\circ \bar{\Phi}\,\,dX\,,
\end{equation}
is nothing but the second variation of $S$ at $\Phi = \bar{\Phi}$. It follows that $B$ is symmetric, as can also be checked explicitly since
\begin{equation}
B_v(\xi,\xi') - B_v(\xi',\xi) = \int_\mathcal{B} ([\mathcal{L}_{\xi},\mathcal{L}_{\xi'}]v) \circ \bar{\Phi}\,dX =
\int_\mathcal{B} (\mathcal{L}_{[\xi,\xi']}v) \circ \bar{\Phi}\,\,dX = 0\,
\end{equation}
where the last equality follows from the Lie algebra property of the $\xi$'s together with (\ref{critical}). The variation principle (\ref{euact}) describes static configurations of the domain $\mathcal{B}$, viewed as a rigid body in the potential $v$.\\
Our work in this paper rests on the (non-)degeneracy properties of the form $B_v$.
 For example when $v$ has continuous symmetries, i.e. $L_\xi v = 0$ for a subalgabra of Euclidean Killing vectors, $B_v$ is degenerate along this subalgebra for any critical $\bar{\Phi}$. The question then is if at least $B_v$ is non-degenerate on the quotient of $\mathbb{E}(3)$ by this subalgebra. \\
Such questions can be fully analyzed in the following cases: Let
$v$ be of the form $v(x) = \frac{1}{2} f_{ij}\, (x^i - x_0^i) (x^j - x_0^j)$ with constants $f_{ij} = f_{(ij)}$\footnote{No confusion between this $f$ and the maps $f^A$ should arise.} and $\bar{\Phi}$ be a map sending $\mathcal{B}$ into $\mathbb{R}^3_\mathcal{S}$ so that
\begin{equation}\label{centre}
D^i \equiv |\mathcal{B}|^{-1}\int_{\mathcal{B}} \bar{\Phi}^i \, dX = x_0^i,
\end{equation}
where $|\mathcal{B}| = \int_{\mathcal{B}} dX$, i.e. the centre of mass is at the critical point of $V$. It now follows that Eq.(\ref{critical}) is valid for translational Killing vectors $\xi^i(x) = c^i$. We are still free to compose $\bar{\Phi}$ with a rotation in $\mathbb{R}_{\mathscr{S}}$. The condition for Eq.(\ref{critical}) to be valid also for rotational Killing vectors is that
\begin{equation}
\int_{\mathcal{B}} \omega^i{}_j \bar{\Phi}^j f_{il}(\bar{\Phi}^l - x_0^l)\,dX = 0
\end{equation}
for arbitrary constants $\omega^i{}_j$ with $\omega_{ij} = \omega_{[ij]}$. This, by (\ref{centre}), is equivalent to the condition that the constant linear map $f^i{}_j$ commutes with $\Theta^i{}_j$, where $\Theta^{ij} = \int_{\mathcal{B}}
(\bar{\Phi}^i - x_0^i)(\bar{\Phi}^j - x_0^j) \,dX$. Now, for any two symmetric matrices $f$ and $\Theta$, it is always possible to find a rotation $R$ so that $R \Theta R^{-1}$ commutes with $f$. Namely let $S$ be a rotation so that, in the given basis, $SfS^{-1}$ is diagonal and $T$ be a rotation so that, in the same given basis, $T \Theta T^{-1}$ is diagonal. Then $SfS^{-1}$ commutes with $T \Theta T^{-1}$. But this is equivalent to that
$R \Sigma R^{-1}$ commutes with $f$
with $R = S^{-1}T$. We call the map $X^A \mapsto R^i{}_j \bar{\Phi}^j(X)$ again $\bar{\Phi}$. Next one can compute
$B_v$. \\
Let $\xi^i(x) = c^i + \omega^i{}_j (x^j - x_0^j) = c^i + \epsilon^i{}_{jk}p^k (x^j - x_0^j)$. One finds that
\begin{equation}\label{H}
B_v(\xi,\xi') = |\mathcal{B}|f_{ij} c^i c'^j + \tau_{ij}p^i p'^j\,,
\end{equation}
where
\begin{equation}
\tau_{ij}p^i p'^j = (f_i{}^k\Theta^l{}_j + f_{mi} \Theta^{ml} \delta^k{}_j)\,\epsilon^i{}_{kr}\, \epsilon^j{}_{ls}\, p^r p\,'^s\,.
\end{equation}
The quantity $\tau$ can in matrix notation be written as
\begin{equation}\label{tau}
\tau = 3 f \Theta - (\mathrm{tr} f) \Theta - (\mathrm{tr} \Theta) f + [(\mathrm{tr} \Theta)(\mathrm{tr} f) - 2\, \mathrm{tr}(f \Theta)] \,\mathrm{id}\,,
\end{equation}
The matrix $\tau$ is symmetric by virtue of $f$ and $\Theta$ commuting. This also means that $\Theta$ whence $\tau$ is simultaneously diagonalizable with $f$. So, denoting the eigenvalues of $f$ resp. $\Theta$ by $(f_1,f_2,f_3)$ resp. $(\Theta_1,\Theta_2,\Theta_3)$, we finally conclude that the signature of $B$ is given by
\begin{equation}
[\epsilon(f_1),\epsilon(f_2),\epsilon(f_3),\epsilon((f_3 - f_2)(\Theta_2 - \Theta_3)),\epsilon( (f_1 - f_3)(\Theta_3 - \Theta_1)),\epsilon((f_2 - f_1)(\Theta_1 - \Theta_2))]\,
\end{equation}
where $\epsilon$ denotes the sign function. Suppose that the $\Theta_i$'s are mutually different. Then each degeneracy in $B$ is due to a continuous symmetry of $v$, namely a rotation, say around the $x^3$-axis when $f_1=f_2$ or a translation, say in the $x^2$-direction when $f_2=0$.


\begin{thebibliography}{99}
\bibitem[1] {B} {Beig, R.,} Concepts of Hyperbolicity and Relativistic Continuum Mechanics, in: Analytical and Numerical Approaches to Mathematical
Relativity  (J. Frauendiener, D.J.W. Giulini, V. Perlick, Eds.) Springer Lecture Notes in Physics, vol.692 101-116 Springer (2006)
\bibitem[2] {BSR} {Beig, R., Schmidt, B.G.,}  Relativistic Elastostatics. I. Bodies in rigid rotation,
{\it Class. Quantum Grav} {\bf 22},
2249-2268 (2005)
\bibitem[3] {BS1} {Beig, R., Schmidt, B.G.,}  Static, self-gravitating elastic bodies, {\it Proc.R.Soc.London}
{\bf A459} 109 -115 (2003)
\bibitem[4] {BS2} {Beig, R., Schmidt, B.G.,} Celestial mechanics of elastic bodies, {\it Math. Z.} {\bf{258}}
381-394 (2008)
\bibitem[5] {BS3} {Beig, R., Schmidt, B.G.,} Relativistic Elasticity, {\it Classical and Quantum Gravity} {\bf{20}}
889-904 (2003)
\bibitem[6] {BS4} {Beig, R., Schmidt, B.G.,} Helical Solutions in Scalar Gravity, {\it Gen. Rel. Grav.} {\bf{41}} 2031-2043 (2009)
\bibitem[7] {Br}  {Broda, S.,}  Helical motion of elastic spheres,  arXiv:1108.0325
    \bibitem[8] {CI} {Ciarlet, P.G.,}  Mathematical Elasticity, Volume 1: Three-Dimensional Elasticity,  North-Holland (1988)
\bibitem[9] {GS} {Golubitsky, M., Schaeffer, D.G.,} Singularities and groups in Bifurcation Theory Vol1, Springer (1985)
\bibitem[10] {MH} {Marsden, J.E., Hughes, T.J.R.,}  Mathematical foundations of elasticity, Dover (1994)
\bibitem[11] {VA} {Valent, T.,}  Boundary Value Problems of Finite Elasticity, Springer (1987)
\end{thebibliography}
\end{document}